\documentclass[10pt,journal,compsoc]{IEEEtran}

\ifCLASSOPTIONcompsoc
  \usepackage[nocompress]{cite}
\else
  \usepackage{cite}
\fi

\ifCLASSINFOpdf
\else
\fi
\usepackage{amsmath,amssymb}
\usepackage{float}
\usepackage{fixltx2e}
\usepackage{stfloats}
\usepackage{url}
\usepackage{tikz}
\usetikzlibrary{shapes.geometric, arrows}
\usepackage{amsmath,amsfonts}
\usepackage{algorithmic}
\usepackage{algorithm}
\usepackage{array}
\usepackage{textcomp}
\usepackage{stfloats}
\usepackage{url}
\usepackage{verbatim}
\usepackage{graphicx}
\usepackage{cite}
\usepackage{multirow}
\usepackage{xcolor}
\usepackage{booktabs}

\usepackage{subcaption}
\usepackage{epsfig}

\usepackage{lscape}
\usepackage{forest}
\tikzset{marrow/.style={midway,red,sloped,fill, minimum height=3cm, single arrow, single arrow
    head extend=.5cm, single arrow head indent=.25cm,xscale=0.3,yscale=0.15,
    allow upside down}}

\usepackage{tikz}

\usepackage{tikz}
  \usetikzlibrary{shapes,arrows,fit,calc,positioning}
  \tikzset{box/.style={draw, diamond, thick, text centered, minimum height=0.5cm, minimum width=1cm}}
  \tikzset{line/.style={draw, thick, -latex'}}
\usetikzlibrary{matrix,shapes}
\begin{document}
%
% paper title
% Titles are generally capitalized except for words such as a, an, and, as,
% at, but, by, for, in, nor, of, on, or, the, to and up, which are usually
% not capitalized unless they are the first or last word of the title.
% Linebreaks \\ can be used within to get better formatting as desired.
% Do not put math or special symbols in the title.
\title{BASiNETEntropy: an alignment-free method for classification of biological sequences through complex networks and entropy maximization}
%
%
% author names and IEEE memberships
% note positions of commas and nonbreaking spaces ( ~ ) LaTeX will not break
% a structure at a ~ so this keeps an author's name from being broken across
% two lines.
% use \thanks{} to gain access to the first footnote area
% a separate \thanks must be used for each paragraph as LaTeX2e's \thanks
% was not built to handle multiple paragraphs
%
%
%\IEEEcompsocitemizethanks is a special \thanks that produces the bulleted
% lists the Computer Society journals use for "first footnote" author
% affiliations. Use \IEEEcompsocthanksitem which works much like \item
% for each affiliation group. When not in compsoc mode,
% \IEEEcompsocitemizethanks becomes like \thanks and
% \IEEEcompsocthanksitem becomes a line break with idention. This
% facilitates dual compilation, although admittedly the differences in the
% desired content of \author between the different types of papers makes a
% one-size-fits-all approach a daunting prospect. For instance, compsoc 
% journal papers have the author affiliations above the "Manuscript
% received ..."  text while in non-compsoc journals this is reversed. Sigh.

\author{Murilo Montanini Breve, Matheus Henrique Pimenta-Zanon and~Fabrício Martins Lopes \\
\IEEEmembership{Universidade Tecnol\'ogica Federal do Paran\'a (UTFPR), Computer Science Department, Alberto Carazzai, 1640, 86300-000, Corn\'elio Proc\'opio, PR, Brazil}% <-this % stops a space
\IEEEcompsocitemizethanks{\IEEEcompsocthanksitem Corresponding Author: Fabr\'icio Martins Lopes\protect\\
% note need leading \protect in front of \\ to get a newline within \thanks as
% \\ is fragile and will error, could use \hfil\break instead.
E-mail: fabricio@utfpr.edu.br}
%\IEEEcompsocthanksitem J. Doe and J. Doe are with Anonymous University.}% <-this % stops an unwanted space
\thanks{}}

% note the % following the last \IEEEmembership and also \thanks - 
% these prevent an unwanted space from occurring between the last author name
% and the end of the author line. i.e., if you had this:
% 
% \author{....lastname \thanks{...} \thanks{...} }
%                     ^------------^------------^----Do not want these spaces!
%
% a space would be appended to the last name and could cause every name on that
% line to be shifted left slightly. This is one of those "LaTeX things". For
% instance, "\textbf{A} \textbf{B}" will typeset as "A B" not "AB". To get
% "AB" then you have to do: "\textbf{A}\textbf{B}"
% \thanks is no different in this regard, so shield the last } of each \thanks
% that ends a line with a % and do not let a space in before the next \thanks.
% Spaces after \IEEEmembership other than the last one are OK (and needed) as
% you are supposed to have spaces between the names. For what it is worth,
% this is a minor point as most people would not even notice if the said evil
% space somehow managed to creep in.

% The paper headers
\markboth{}%
{Breve \MakeLowercase{\textit{et al.}}: Bare Demo of IEEEtran.cls for Computer Society Journals}
% The only time the second header will appear is for the odd numbered pages
% after the title page when using the twoside option.
% 
% *** Note that you probably will NOT want to include the author's ***
% *** name in the headers of peer review papers.                   ***
% You can use \ifCLASSOPTIONpeerreview for conditional compilation here if
% you desire.

% for Computer Society papers, we must declare the abstract and index terms
% PRIOR to the title within the \IEEEtitleabstractindextext IEEEtran
% command as these need to go into the title area created by \maketitle.
% As a general rule, do not put math, special symbols or citations
% in the abstract or keywords.
\IEEEtitleabstractindextext{%
\begin{abstract}
The discovery of nucleic acids and the structure of DNA have brought considerable advances in the understanding of life. The development of next-generation sequencing technologies has led to a large-scale generation of data, for which computational methods have become essential for analysis and knowledge discovery. In particular, RNAs have received much attention because of the diversity of their functionalities in the organism and the discoveries of different classes, such as mRNA, tRNA, ncRNA, sncRNA, lncRNA, among others, which have different functions in many biological processes. Therefore, the correct identification of RNA sequences from unknown biological sequences is increasingly important to provide relevant information to understand the functioning of organisms. This work addresses this context by presenting a new method for the classification of biological sequences through complex networks and entropy maximization. The proposed method initially maps the RNA sequences and represents them as complex networks. Then, an approach based on the maximum entropy principle is proposed to identify the most informative edges about the RNA class. Then, the most informative edges are kept in the complex network and the less informative ones are removed, generating a filtered complex network. This filtered network is then characterized through topological measures, which are used in the classification process. The proposed method was evaluated in the classification of mRNAs, ncRNAs, sncRNAs and lncRNAs from 13 species. The proposed method was compared to PLEK, CPC2 and BASiNET methods, outperforming all compared methods. BASiNETEntropy classified all RNA sequences with high accuracy and low standard deviation in results, showing that the method is robust and not biased by the organism. The proposed method is implemented in an open source in R language and is freely available for download at \url{https://cran.r-project.org/web/packages/BASiNETEntropy}.
\end{abstract}

% Note that keywords are not normally used for peerreview papers.
}

% make the title area
\maketitle

% To allow for easy dual compilation without having to reenter the
% abstract/keywords data, the \IEEEtitleabstractindextext text will
% not be used in maketitle, but will appear (i.e., to be "transported")
% here as \IEEEdisplaynontitleabstractindextext when the compsoc 
% or transmag modes are not selected <OR> if conference mode is selected 
% - because all conference papers position the abstract like regular
% papers do.
\IEEEdisplaynontitleabstractindextext
% \IEEEdisplaynontitleabstractindextext has no effect when using
% compsoc or transmag under a non-conference mode.

% For peer review papers, you can put extra information on the cover
% page as needed:
% \ifCLASSOPTIONpeerreview
% \begin{center} \bfseries EDICS Category: 3-BBND \end{center}
% \fi
%
% For peerreview papers, this IEEEtran command inserts a page break and
% creates the second title. It will be ignored for other modes.
\IEEEpeerreviewmaketitle

% **********************************************************************
\section{Introduction}%
\label{sect:introduction}

In the 19th century, the Swiss biochemist Friedrich Miescher (1844-1895) isolated an acid containing phosphorus and nitrogen from a cell. Twenty years after his discovery, Richard Altmann named this compound nucleic acid, as we know it today \cite{DAHM2005274}. Because of these scientific advances, in 1953, James D. Watson and Francis H. Crick published ``A Structure for Deoxyribose Nucleic Acid'' \cite{FEUGHELMAN1955}, the first mention of DNA structure in the scientific field, and it would be one of the most important contributions to biology.

Because of its indispensable participation in the creation and maintenance of life, nucleic acids have gained space and notoriety in various fields of knowledge, such as in computer science, since the amount of information present in the agglomerates of these particles is enormous, which makes it impractical to analyze manually. Since Phage-X174 was sequenced in 1977, a huge amount of organisms have been sequenced and stored in databases. In this context, the need for a computer application in biology created a new term, Bioinformatics \cite{hogeweg_roots_2011}, aiming to define the study of computational processes in biotic systems. Since then, bioinformatics methods have become essential for analysis of biological data \cite{posada2009bioinformatics}. Nowadays, computer programs such as NCBI BLAST are routinely used to perform a comparative analysis of biological sequences over 9.9 trillion base pairs and 2.1 billion nucleotide sequences \cite{GenBank2020}.

Regarding genetics and genomics, bioinformatics methods are mainly involved in the sequencing and annotation of an organism’s DNA assemblies, and their observed mutations \cite{hagen_origins_2000,lander_initial_2011,hogeweg_roots_2011}. DNA is composed by nucleic acids that stores genetic information by combining four types of nitrogenous bases (adenine (A), thymine (T), guanine (G) and cytosine (C)), which will form distinct DNA molecules according to the structural organization of its bases. The information in the genome largely defines the functionality that a gene exerts in the organism \cite{Varki}, hereditary characteristics are also passed on through the genome \cite{Wolf}.

On the other hand, RNA is the other type of nucleic acid composed of four different nucleotide subunits joined by phosphodiester bonds, and one of its functions is to serve as a template for proteins production (translation). RNA sequences are analysed to determine which genes code proteins and also to compare genes within a species or between different species, which can show similarities in protein functions or relationships between species. There are several classes of RNA, such as messenger RNAs (mRNAs) that are protein-coding, ribosomal RNAs (rRNA), transporter RNAs (tRNAs), non-coding RNAs (ncRNAs) among others\cite{alberts2017molecular}.
In particular, the ncRNA recently received much attention because of its functionalities and diversity, such as small non-coding (sncRNA) and long non-coding (lncRNA) and their respective subclasses \cite{amin2019evaluation}.

Long non-coding RNAs (lncRNAs) are emerging as an important component in the cancer context, indicating potential roles in oncogenic and tumour suppressor pathways \cite{Gib}. Small non-coding RNAs (sncRNAs) are part of non-coding regulatory oligo-nucleotides with broad physiological and morphological functions. They control the genetic programming of cells and can modulate differentiation and death processes\cite{KLIMENKO201788}.

In fact, the ncRNAs are still little known, however it is already possible to point out important functionalities in many biological processes, transcriptional regulation of gene expression, mediate epigenetic modifications of DNA and also their relationships in cancer and other human diseases, such as neurological, cardiovascular and autoimmune disorders \cite{Esteller2011, Rinn2012, dai2020rna}. 

Therefore, the correct identification of RNA classes between mRNA, lncRNA and sncRNA from unknown biological sequences may contribute to a better understanding of their functionalities and mechanisms. This work addresses this context by presenting a new method for the classification of biological sequences through complex networks and entropy maximization.

% **********************************************************************
\section{Biological Background}
\label{sec:biological}

\begin{figure}[!h]
\centering%
\includegraphics[width=1\linewidth]{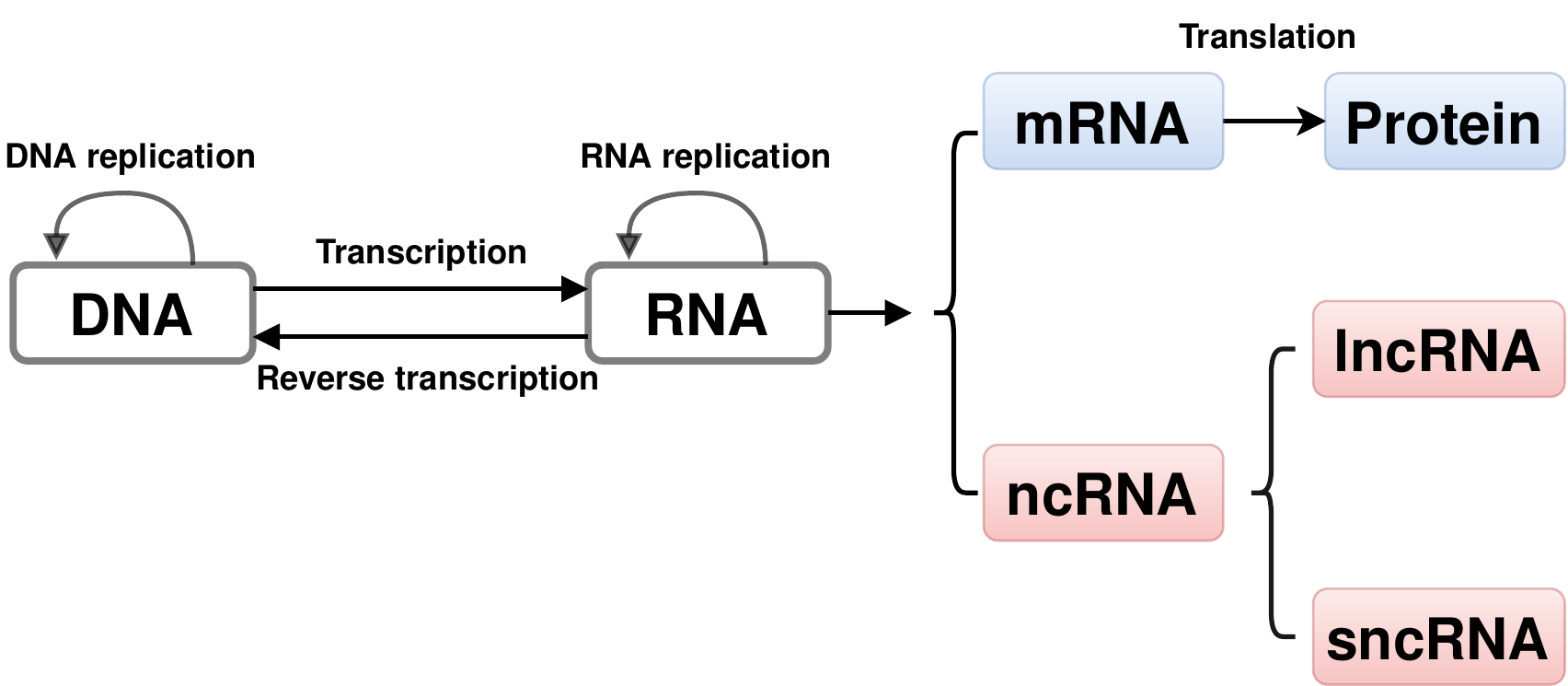}
\caption{Expanded central dogma of molecular biology.}%
\label{fig:dogma}
\end{figure}

The central dogma of molecular biology reveals the structure of DNA and explains the flow of genetic information, from DNA to RNA, to make a functional product, a protein \cite{crick1970central}. Thus, the information in DNA is transcribed to synthesise RNA molecules. RNAs that encode proteins are defined as messenger RNA (mRNA) and are widely studied given their protein-coding function. On the other hand, about 1.5\% of the human genome is transcribed into mRNA, thus a great part of the transcriptome does not possess the ability to code proteins, which are defined as non-coding RNA (ncRNA) \cite{zheng_systematic_2021}. In generality, non-coding RNA can be classified given the size of the transcript, being classified into small ($<$ 200 nucleotides) (sncRNA) and long ($>$ 200 nucleotides) (lncRNA) \cite{qin_structure_2020}. Figure \ref{fig:dogma} presents an overview of the expanded central dogma of molecular biology.

A large proportion of non-coding RNA is present in the nucleus of cells, when compared with mRNA. In particular, a large proportion of lncRNA are transcribed from RNA polymerase II, showing a transcription mechanism similar to the mRNA \cite{statello_gene_2021}. Unlike lncRNA, sncRNAs are generated at various stages of transcription, and may be a product of the separation of introns, in the generation of inhibitors and other stages of transcription \cite{Romano2017}.

sncRNA act in many developmental processes, including cell maintenance, development and differentiation, transcriptional and post-transcriptional gene silencing \cite{ncRNA2010}. The dysregulation of some sncRNA is associated with several diseases such as cancer, neurological and cardiac diseases among others \cite{Esteller2011, Chen2016}, thus sncRNAs are used as biomarkers for the detection of some of these diseases \cite{Romano2017}. 

The function of lncRNAs can be classified into three major groups based on their localization in the cell, being a regulation of chromatin states and gene expression, influence on nuclear structure, and regulation of proteins and other RNA molecules \cite{zheng_systematic_2021}. Besides acting in regulation, some diseases are related to lncRNA, such as hematopoiesis, some types of cancer, and diseases related to immune and neurological responses \cite{statello_gene_2021}.

% **********************************************************************
\section{Related Works}
\label{sec:relatedworks}
Traditionally, biological sequence analysis is performed using alignment tools and even using computational techniques, such as dynamic programming and optimization methods, has a high computational cost, making its application in large volumes of data unfeasible \cite{de_pierri_sweep_2020}. Several computational tools using the alignment approach were proposed in the last years to solve RNAs prediction and classification such as Coding Potential Calculator (CPC)~\cite{CPC2007}, LncRNA-ID~\cite{LncRNA-ID2015} and PLncPRO~\cite{PLncPRO2017}.

To overcome this limitation, alignment-free methods have been proposed, and these methods combine data mining, pattern recognition, and machine learning techniques to solve biological problems. In particular, for RNA classification different alignment-free approaches have been proposed \cite{zheng_systematic_2021}.

Considering the context of RNA classification using the alignment-free approach, some approaches have been proposed in the literature. Among the main approaches available in the literature are PLEK \cite{PLEK2014}, CPC2 \cite{CPC22017}, and BASiNET \cite{BASiNET2018}. 

PLEK (\textit{predictor of long non-coding RNAs and messenger RNAs based on an improved k-mer scheme}) \cite{PLEK2014} is an alignment-free method aiming to classify the RNA sequences into long non-coding and coding RNA. Assuming a k-mer frequency ranging from 1 to 5, features are extracted from the raw sequence using a sliding window with a step of one nucleotide. For each k value, the pattern number increases, i.e., for 1-mer have 4  patterns, for 2-mer have 16 patterns, until 1,024 patterns for 5-mer. Each k-mer pattern is incremented using the frequency of occurrence. These frequencies are calibrated and used as features in the support vector machine (SVM) algorithm to build a binary classification model to identify lncRNAs from mRNAs. 

Using a feature list, the CPC2 method \cite{CPC22017} identifies the effective features employing the recursive feature elimination (RFE) approach, and the Fickett TESTCODE score, open reading frame (ORF) length, ORF integrity, and isoelectric point are used as features in the SVM algorithm to classify the coding and non-coding RNA. To create the feature list, the existence of known protein sequences is necessary, i.e., the feature extraction is dependent on data other than the nucleotides sequences. As the CPC2 method depends on known sequences, it has a limitation regarding to extract features from \textit{de novo} sequencing of new organisms or proteins.

More recently, BASiNET \cite{BASiNET2018} was proposed as an alignment-free approach to classify biological sequences based on the use of complex networks and thresholds. The method is based on the mapping of a biological sequence to a weighted complex network, from which topological measurements are extracted for its characterisation \cite{costa_characterization_2007}. Then, thresholds are applied iteratively to remove the less frequent edges and to produce topological measurements at each application of the thresholds. The thresholds are applied until there are edges (maximum 200 iterations). Ten topological measures are extracted at each iteration, thus producing up to 2000 different measurements for each biological sequence. The produced complex network measurements are merged into a single feature vector, one for each biological sequence, and it is adopted as the input for the classification of RNA sequences. However, this huge amount of features leads to high dimensionality in the feature space and increases the computational complexity in terms of processing time and memory.

% **********************************************************************
\section{Complex Networks}
\label{sec:complexnetworks}

Complex networks are graphs with non-trivial topologies, which have a set of vertices (nodes) that are connected through edges \cite{Barabasi}. In fact, many interactions in the real world have connections that can be represented for complex networks, such as social relationships \cite{silva2022}, eletric power grid \cite{albert2004structural}, internet \cite{maslov2004detection}, computer vision \cite{backes2013,lima2015,piotto2016,lima2019,piotto2021} and biological systems \cite{lopes2011b,lopes2014a,Montanini2020}, among others.

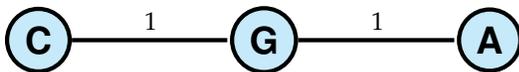
\begin{figure}[h]
\centering%
\begin{tikzpicture}[shorten >=1pt, auto, node distance=3cm, ultra thick]
   \begin{scope}[every node/.style={circle,draw=black,fill=cyan!20!,font=\sffamily\Large\bfseries}]
    \node (v1) at (0,0) {C};
    \node (v2) at (3,0) {G};
    \node (v3) at (6,0) {A};

   \end{scope}
   \begin{scope}[every edge/.style={draw=black,ultra thick}]
    \draw  (v1) edge node{1} (v2);
    \draw  (v2) edge node{1} (v3);

   \end{scope}
\end{tikzpicture}
\caption{An example of weighted graph generated from the sequence: CGA.}%
\label{grafo}
\end{figure}

A graph is composed of a set of nodes (or vertices) and a set of edges that represent the connections among the nodes. Depending on the application, edges may have a direction and an associated weight.
When there is an edge between two nodes, they are adjacent. Figure \ref{grafo} presents and example of graph, in which the nodes C and A are not adjacent, however, the nodes C and G, and G and A are adjacent. A graph is usually represented by its adjacency matrix, as shown in Table \ref{adj}, which is a n-by-n matrix whose value in row i and column j gives the weight of edges from the i-th to the j-th nodes \cite{costa_characterization_2007}.

\begin{table}[h]
\centering
\caption{Adjacency matrix of the graph (Figure \ref{grafo}).}
\label{adj}
\begin{tabular}{c|ccll}
\textit{}  & \textit{\textbf{A}} & \textit{\textbf{C}} & \textbf{T} & \textbf{G} \\ \hline
\textbf{A} & 0                   & 0                 & 0        & 1  \\
\textbf{C} & 0                   & 0                 & 0        & 1  \\
\textbf{T} & 0                   & 0                 & 0        & 0  \\
\textbf{G} & 1                   & 1                 & 0        & 0
\end{tabular}
\end{table}

The complex network theory presents well-defined topologies that describe the structure and dynamics of an network. In face of this, it is possible to extract measurements to characterize its topological structure \cite{BOCCALETTI2006175,costa_characterization_2007}. For the representation of the topology, in this work were adopted 10 measures commonly used in the current literature for the characterization of complex networks. The adopted measurements are briefly described below.

\begin{itemize}
     \item  Average shortest path length (ASPL): is the average of all the minimum paths in a complex network. \cite{BOCCALETTI2006175}. Its value depends on how concentrated the complex network is, i.e., it is low for very concentrated networks.
     \item Clustering coefficient (CC): has the purpose of calculating the probability that nodes are connected to another node that are also connected to each other. For example, applying this measure to a social network would have the effect of calculating the probability that two people who know a third person, which also know each other. Transitivity can be seen as the formation of triangles in complex networks \cite{BOCCALETTI2006175}.
     \item Maximum degree (MAX): is the measure that presents the node with the largest number of connected edges.
     \item Minimum degree (MIN): is the measure that presents the node with the smallest number of connected edges.
     \item Average degree (DEG): is the average number of connections of the network nodes.
     \item Assortativity (ASS): is the probability that nodes with similar degrees are connected. Positive assortativity values mean that nodes of similar degree tend to connect and negative assortativity values mean the opposite \cite{Panwar2014}.
     \item Average standard deviation (ASD): is the standard deviation of the nodes degree, high values of the average standard deviation show that network has unbalanced numbers of connections, i.e., there are nodes that connect more than others.
     \item Average betweenness centrality (BET): is a standard measure of node centrality that show how relevant a node is by considering the number of shortest paths going through it \cite{costa_characterization_2007,BOCCALETTI2006175}. 
       \item  Frequency of motifs with size 3 (MT3): motifs are subgraphs or patterns with various shapes that exist within a network. This measure presents the count of how many motifs of size 3 occur within a network.
       \item  Frequency of motifs with size 4 (MT4): similar to previous measure, is the count of how many motifs of size 4 occur within a network.
   \end{itemize}

%*************************************************************************************************
\section{Entropy}
\label{sec:entropy}
The concept of entropy was introduced in 1865 by Rudolf Clausius in thermodynamics, considering only macroscopic demonstrations \cite{clausius1879}. A few years later, in 1877, Ludwig Boltzmann showed that entropy can be expressed in terms of probabilities associated with the microscopic configuration of a system \cite{boltzmann1974}, which came to be known in the literature as Boltzmann-Gibbs entropy. Later in 1948, entropy was applied to Information Theory by Claude Shannon \cite{shannon1948} and also often called Shannon's entropy. Entropy is often used to indicate the amount of information in a given source, and is also used to measure the disorder (uncertainty) of a data set \cite{bishop1995}. Consider a random variable $X$ that can take on a discrete value. The Shannon entropy \cite{shannon1963}, like the Boltzmann-Gibbs entropy, is defined in terms of the probabilities of the possible occurrences of this random variable $P(x)$, as follows:

\begin{equation}
    \label{eq:shannonentropy}
    H(X) = -\sum_{x \in X} \ P(x) \log P(x) \textrm{,}
\end{equation}

\noindent so that

\begin{eqnarray}
     \nonumber
     \sum_{x \in X} P(x) = 1 \textrm{.}
\end{eqnarray}

In Equation~\ref{eq:shannonentropy} the average of the logarithms of the probabilities of the occurrences $x$ ($log(P(x)$) weighted by their probabilities $P(x)$ is taken, being assumed $0 \times log(0) = 0$. Thus, entropy represents a measure of uncertainty associated with a variable, i.e., the higher the entropy of a variable, the greater the uncertainty in predicting the value of that variable.

Since then, entropy has been used in various fields of knowledge, from classical thermodynamics, where it was first proposed, to statistical physics and the information theory. Over time, this term, found far-reaching applications in chemistry and physics, and currently the entropy also takes part in the study of biological systems and their relationship to life. Several methods with different application using the entropy concept emerged. One of the most important was the development of the maximum entropy (ME) method by the physicist Edwin Thompson Jaynes \cite{Ja}. It was showed that statistically maximizing entropy to observe how gas molecules were distributed would be equivalent to simply maximizing Shannon’s entropy \cite{6773024, Ja}. In fact, the ME principle can be applied to measure the amount of uncertainty contained in a probability distribution \cite{guiasu1985principle}.

Let $h_1,h_2,\ldots,h_n$ be the observed frequencies of a discrete distribution and let
\begin{equation}
    p_i = \frac{h_i}{N}, \quad \sum_{i=1}^{n} h_i = N, \quad i=1,2,\ldots,n,
    \label{eq:frequencies}
\end{equation}
\noindent where $N$ is the total number of samples, $n$ is the number of events (possible outcomes or states of a system) and $p_i$ is the probability of the $i$-$th$ outcome.

Considering a discrete distribution containing two classes $A$ and $B$, its respective entropy can be defined as follows:
\begin{equation}
    H(A) = - \sum_{i=1}^{s} \frac{p_i}{P_A} \log \frac{p_i}{P_A},
    \label{eq:MEA}
\end{equation}

\begin{equation}
    H(B) = - \sum_{i=s+1}^{n} \frac{p_i}{P_B} \log \frac{p_i}{P_B},
    \label{eq:MEB}
\end{equation}

\begin{equation}
    P_A = \sum_{i=1}^{s} p_i, \quad  P_B = \sum_{i=s+1}^{n} p_i, \quad P_A + P_B = 1.
\end{equation}

Given the entropy of each class $A$ and $B$ its is possible to build the distribution of the sum of entropies $H(A)+H(B)$, then ME can be defined as follows:
\begin{equation}\label{eq:3}
%ME = argmax  H_A + H_B, \\
ME = \mathop{\arg \max}\limits_{s = 1,2,\ldots,n} \{H(A)+H(B)\}.
\end{equation}

In this context, its is possible to identify the separability of a system regarding its probability distribution, i.e., to find $s$ into the distribution that produces the maximum ME, leading to the maximum uncertainty between the classes $A$ and $B$ \cite{guiasu1985principle}.

Many applications were proposed over time in a wide variety of scientific research based on ME principle \cite{kapur1992entropy,banavar2010applications,Kumar2021}. 
In particular, there are bioinformatics and computational biology methods based on ME principle \cite{Morcos2011, granot2013stimulus, Wouter2014, barros_2017, bottaro2020integrating}, which have proven very suitable and becoming increasingly useful.

% *************************************************************************************************
\section{Materials and Methods}

\subsection{Materials}
\label{sec:materials}
In order to assess the proposed method, two datasets were adopted. These datasets were commonly used and also allow the results to be compared openly with other methods. The first dataset was obtained from PLEK \cite{PLEK2014} and the second dataset was obtained from CPC2 \cite{CPC22017}. These two datasets contain different RNAs classes from different species.

The specification of species, classes and the number of sequences per organisms is showed in Tables \ref{tab:plek1} and \ref{tab:cpc2}. Regarding the RNA classes, two are available in PLEK dataset:  messenger RNA (mRNA) and long non-coding RNA (lncRNA) and three are available in CPC2 dataset: mRNA, lncRNA and small non-coding RNA (sncRNA). The datasets contain 13 different species, including vertebrates, plant, worm and insect. There are three shared species between the two data sets: \textit{Homo sapiens}, \textit{Mus musculus} and \textit{Danio rerio}. 

\begin{table}[!h]
\centering%
\caption{PLEK dataset \cite{PLEK2014}.}
\label{tab:plek1}
\begin{tabular}{|c|c|r|}
\hline
Species                    & Class of RNA & Sequences \\ \hline
            
\multirow{2}{*}{\textit{Homo Sapiens}}       & mRNA                 & 4127                           \\ \cline{2-3} 
                                    & ncRNA                & 22389                            \\ \hline   
\multirow{2}{*}{\textit{Mus musculus}}       & mRNA                 & 26062                          \\ \cline{2-3} 
                                    & ncRNA                & 2963                           \\ \hline
\multirow{2}{*}{\textit{Danio rerio}}        & mRNA                 & 14493                          \\ \cline{2-3} 
                                    & ncRNA                & 419                            \\ \hline
\multirow{2}{*}{\textit{Bos taurus}}         & mRNA                 & 13190                          \\ \cline{2-3} 
                                    & ncRNA                & 182                            \\ \hline
\multirow{2}{*}{\textit{Gorilla gorilla}}    & mRNA                 & 33025                          \\ \cline{2-3} 
                                    & ncRNA                & 367                            \\ \hline
\multirow{2}{*}{\textit{Macaca mulatta}}     & mRNA                 & 5709                           \\ \cline{2-3} 
                                    & ncRNA                & 359                            \\ \hline
\multirow{2}{*}{\textit{Pan troglodytes}}    & mRNA                 & 1906                           \\ \cline{2-3} 
                                    & ncRNA                & 1166                           \\ \hline
\multirow{2}{*}{\textit{Pongo abelii}}       & mRNA                 & 3401                           \\ \cline{2-3} 
                                    & ncRNA                & 392                            \\ \hline
\multirow{2}{*}{\textit{Sus scrofa}}         & mRNA                 & 3978                           \\ \cline{2-3} 
                                    & ncRNA                & 241                            \\ \hline
\multirow{2}{*}{\textit{Xenopus tropicalis}} & mRNA                 & 8874                           \\ \cline{2-3} 
                                    & ncRNA                & 279                            \\ \hline
\end{tabular}
\end{table}

\begin{table}[!h]
\centering%
\caption{CPC2 dataset \cite{CPC22017}.}
\label{tab:cpc2}
\begin{tabular}{|c|c|r|}
\hline
Species                                                                                      & Class of RNA & \multicolumn{1}{c|}{Sequences} \\ \hline
\multirow{3}{*}{\textit{Homo sapiens}}                                                       & mRNA         & 6142                           \\ \cline{2-3} 
                                                                                             & lncRNA       & 7485                           \\ \cline{2-3} 
                                                                                             & sncRNA       & 4534                           \\ \hline
\multirow{3}{*}{\textit{Mus musculus}}                                                       & mRNA         & 10638                          \\ \cline{2-3} 
                                                                                             & lncRNA       & 6460                           \\ \cline{2-3} 
                                                                                             & sncRNA       & 5791                           \\ \hline
\multirow{3}{*}{\textit{Danio rerio}}                                                        & mRNA         & 2344                           \\ \cline{2-3} 
                                                                                             & lncRNA       & 1163                           \\ \cline{2-3} 
                                                                                             & sncRNA       & 365                            \\ \hline
\multirow{3}{*}{\textit{\begin{tabular}[c]{@{}c@{}}Arabidopsis \\ thaliana\end{tabular}}}    & mRNA         & 13986                          \\ \cline{2-3} 
                                                                                             & lncRNA       & 2562                           \\ \cline{2-3} 
                                                                                             & sncRNA       & 1291                           \\ \hline
\multirow{3}{*}{\textit{\begin{tabular}[c]{@{}c@{}}Caenorhabditis \\ elegans\end{tabular}}}  & mRNA         & 3551                           \\ \cline{2-3} 
                                                                                             & lncRNA       & 1582                           \\ \cline{2-3} 
                                                                                             & sncRNA       & 7888                           \\ \hline           
\multirow{3}{*}{\textit{\begin{tabular}[c]{@{}c@{}}Drosophila \\ melanogaster\end{tabular}}} & mRNA         & 3680                           \\ \cline{2-3} 
                                                                                             & lncRNA       & 2776                           \\ \cline{2-3} 
                                                                                             & sncRNA       & 780                            \\ \hline

\end{tabular}
\end{table}

% ********************************************************************************************************
\subsection{Methods}
\label{sec:methods}
The development of this work is based on the BASiNET \cite{BASiNET2018} method. In this way, graphs (complex networks) were adopted, which were generated from the RNA sequences from PLEK \cite{PLEK2014} and CPC2 \cite{CPC22017} datasets.

\begin{figure}[h]
\centering%
\includegraphics[width=1\linewidth]{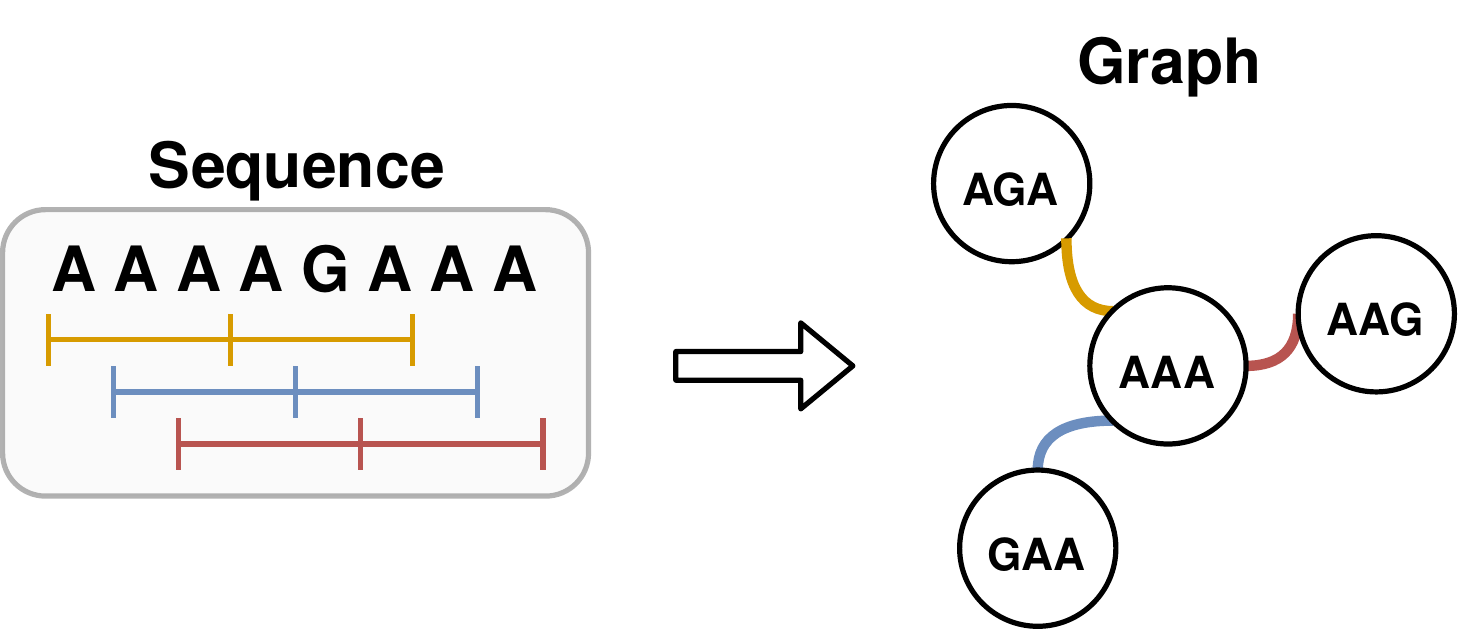}
\caption{Overview of mapping RNA sequences onto graphs.}%
\label{fig:graph1}
\end{figure}

To produce these complex networks, it was necessary to set two parameters, the Step and the Word sizes. The function of the Step size parameter is to define the distance that will be travelled in the sequence after an edge has been formed. Meanwhile, the Word size parameter refers to how many nucleotides will be considered at each node. This process is presented in Figure \ref{fig:graph1}. In this work, similar to the BASiNET, the $step = 1$ and $word = 3$ were adopted.

With the graphs produced, it is proposed an approach to select the most informational edges from the RNA class. Based on maximum entropy principle (Sec. \ref{sec:entropy}), the edges of all networks produced for each RNA class were appended composing a single network. Then, by considering the edges frequency (i.e., weights), a histogram was produced. Thus, similarly that image thresholding method \cite{Kapur}, the aim is to find the threshold ($s$) that maximizes the sum of the entropies of two distinct parts (informational and non-informational) edges of each RNA class. In this way, a filter for the network nodes is proposed, by considering the 4096 possibilities (edges), resulting from a 64x64 matrix (word size = 3), to identify what edges are important and what are not important for each RNA class.

\begin{figure}[h]
\centering%
\includegraphics[width=1\linewidth]{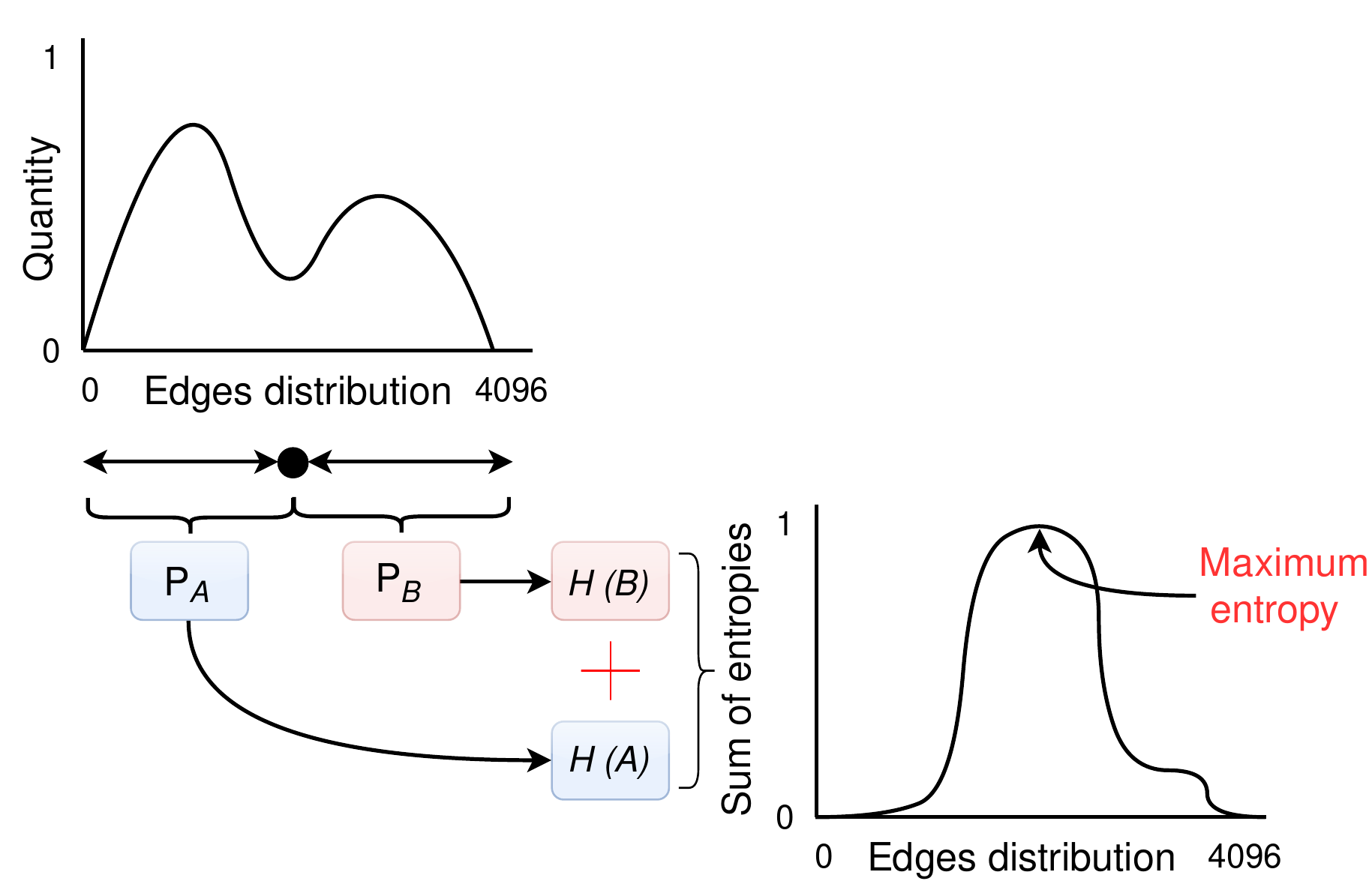}
\caption{Overview of the building $H(A)+H(B)$ entropy distribution and its maximum entropy $ME$.}%
\label{fig:histo}
\end{figure}

The first step of the modelling is to estimate the probabilities of each class $A$ and $B$, i.e., $P_A$ and $P_B$. Then the histogram is traversed from $i=1, \ldots, s$ to estimate $P_A$ and from $s+1, \ldots, n$ to estimate $P_B$, in an iterative way $s=1,2,\ldots,n$ (Sec. \ref{sec:entropy}). Then, the entropy $H(A)$ and $H(B)$ can be also estimated to build the respective $H(A)+H(B)$ distribution and to find its maximum entropy $ME$, i.e., the threshold to identify the informational and non-informational edges of a RNA class. Figure \ref{fig:histo} presents an overview of the proposed approach.

\begin{figure}[h]
\centering%
\includegraphics[width=1\linewidth]{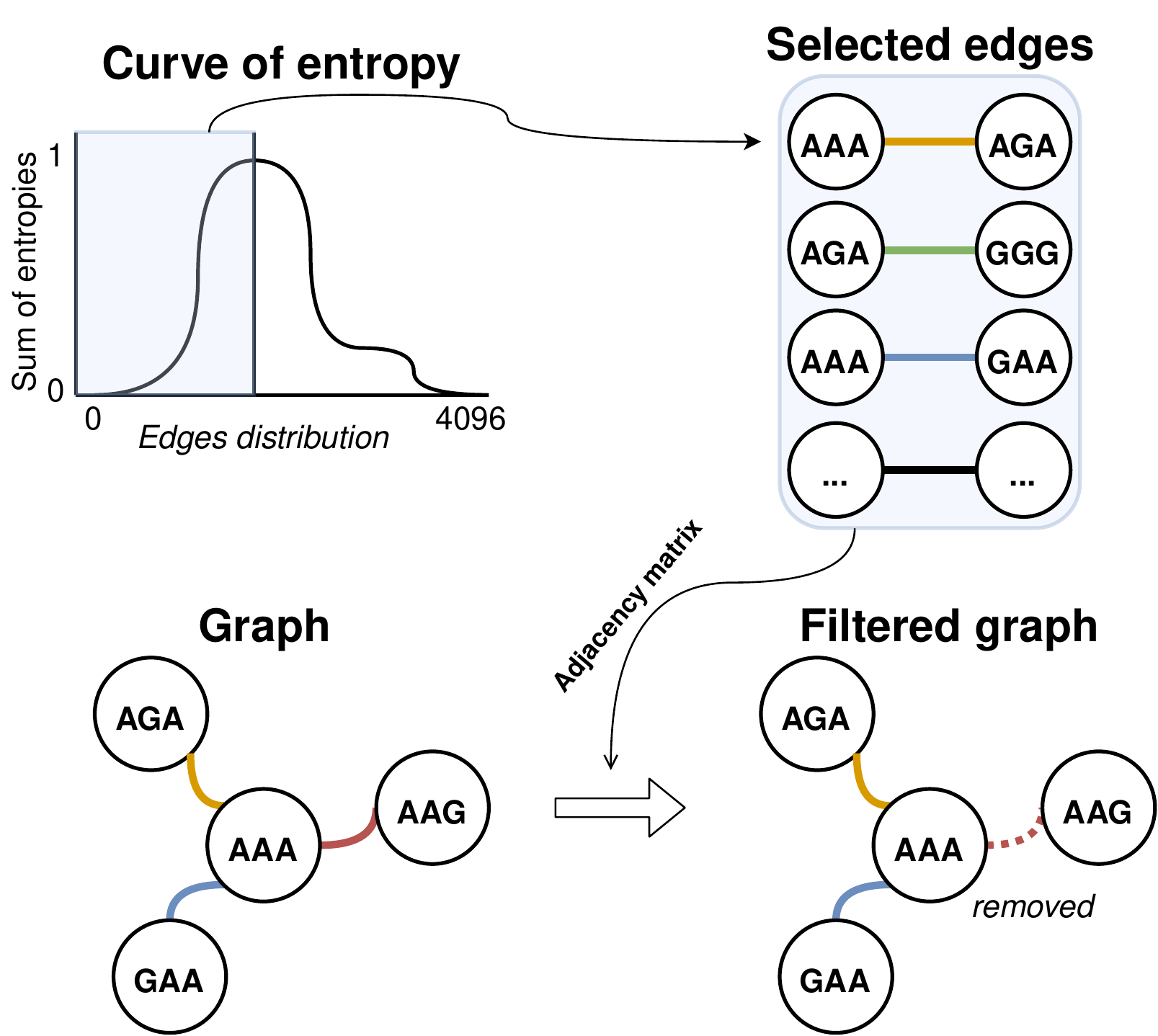}
\caption{Overview of the proposed edge filtering through maximum entropy.}%
\label{fig:filtering}
\end{figure}

The selected edges as informational by the maximum entropy are considered for each RNA class and the non-informational edges are removed, producing a filtered complex network for each class. Figure~\ref{fig:filtering} shows the filtering process. 

\begin{figure}[h!]
\centering%
\includegraphics[width=1\linewidth]{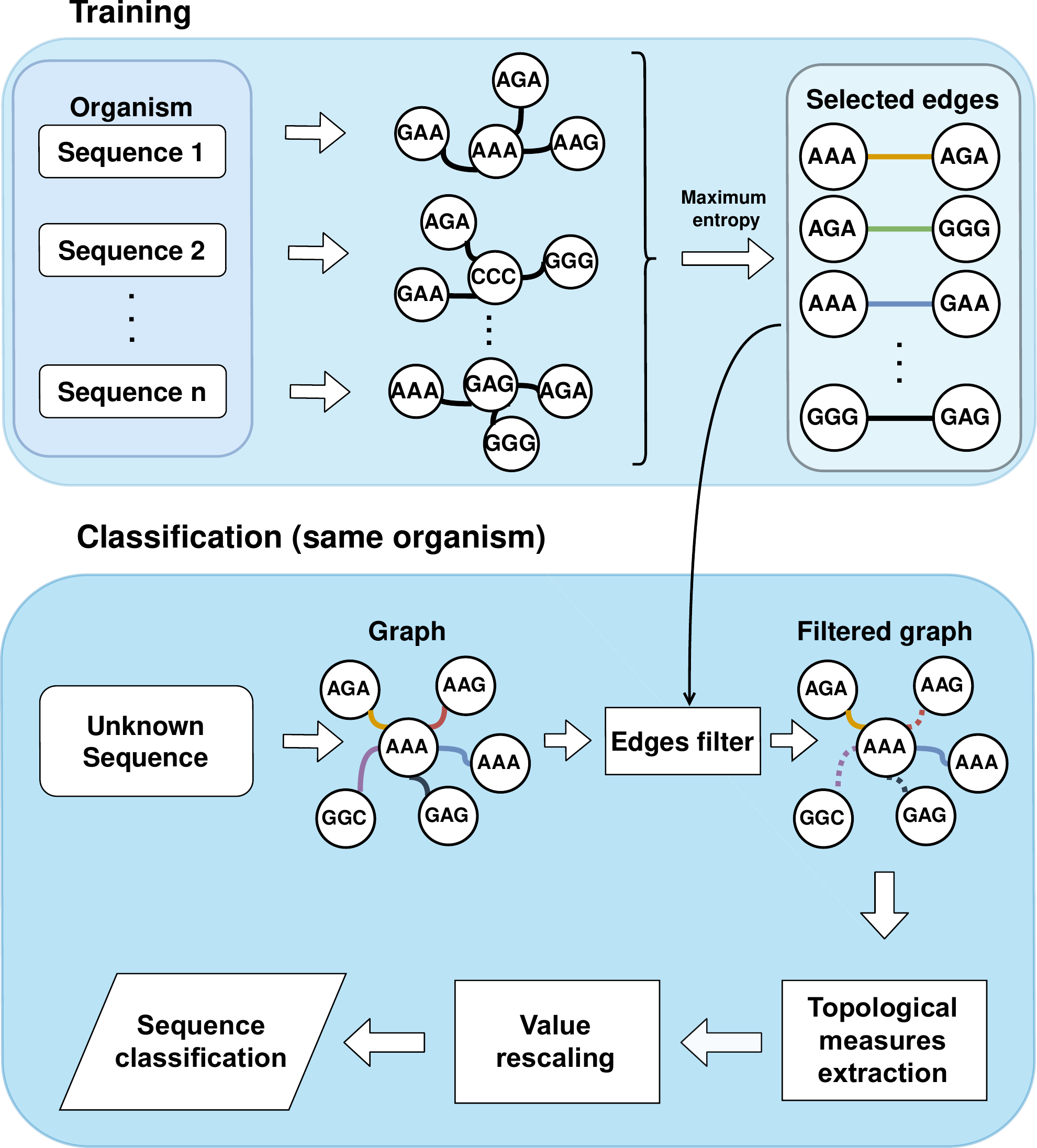}
\caption{Overview of the proposed method and its steps.}
\label{fig:diagram}
\end{figure}

Thus, a filtered complex network is produced for each RNA class and the adopted measurements are extracted for its characterization (Sec. \ref{sec:complexnetworks}). Considering that each complex network measurement has a different numerical interval, for example, the average shortest path length measure is usually found on the tens scale. Meanwhile, other measures can reach values in the hundreds or even thousands scale, which can make some measures more relevant than others to the classifier. Therefore, it is necessary to pre-process the adjust their values so that they are comparable to each other. In this work, a scaling factor is adopted, commonly known as Min-Max, which scales the values of the measures in the interval between 0 and 1. The Equation \ref{eq:et} defines the rescale of the values.

\begin{equation}\label{eq:et}
V{ri} = \frac{V_i-Min_m}{Max_m - Min_m}, \quad i \in N
\end{equation}
\noindent where $N$ is the total number of samples, $V_i$ is the i-th measure value, $Min_m$ is the minimum value of the measure and $Max_m$ is the maximum value of the measure.

It is important to highlight that the proposed method is based on the identification of the most informational edges of each class, a training is required for each of the classes involved. Figure \ref{fig:diagram} presents the overview of the proposed method.

%***************************************************************************************
\section{Results and Discussion}
In order to evaluate the proposed method, two datasets presented in Sec. \ref{sec:materials} were considered. The results were compared with important methods such as: PLEK \cite{PLEK2014}, CPC2 \cite{CPC22017}, BASiNET \cite{BASiNET2018} and also with BASiNET* without considering its iterative threshold step, i.e., producing the same amount of features as the proposed method. In addition, since the adopted datasets are used in other work available in the literature, the experimental results can be directly compared.

\begingroup
\setlength{\tabcolsep}{12pt} % Default value: 6pt
\renewcommand{\arraystretch}{1.1} % Default value: 1
\begin{table*}[h]
\centering
\caption{Classification results considering the mRNA and ncRNA classes of sequences from PLEK dataset~\cite{PLEK2014}.}
\label{tab:plek}
\begin{tabular}{ccccccc} 
\hline
\begin{tabular}[c]{@{}c@{}}Species\end{tabular} & Class of RNA & PLEK           & CPC2           & BASiNET*       & BASiNET        & BASiNETEntropy  \\ 
\hline
\multirow{2}{*}{\textit{Homo Sapiens}}            & mRNA         & 96.7           & 94.3           & 54.4           & 99.9           & 99.6            \\
                                                  & ncRNA        & 99.3           & 93.9           & 95.6           & 100.0          & 100.0           \\ \hline
\multirow{2}{*}{\textit{Mus musculus}}            & mRNA         & 88.1           & 94.7           & 79.9           & 100.0          & 99.9            \\
                                                  & ncRNA        & 89.9           & 99.9           & 75.9           & 99.9           & 100.0           \\\hline
\multirow{2}{*}{\textit{Danio rerio}}             & mRNA         & 91.3           & 96.6           & 99.8           & 100.0          & 100.0           \\
                                                  & ncRNA        & 90.9           & 94.0           & 47.9           & 98.9           & 99.5            \\\hline
\multirow{2}{*}{\textit{Bos taurus}}              & mRNA         & 94.8           & 95.9           & 91.2           & 100.0          & 100.0           \\
                                                  & ncRNA        & 99.5           & 100.0          & 99.8           & 98.9           & 99.5            \\\hline
\multirow{2}{*}{\textit{Gorilla gorilla}}         & mRNA         & 83.8           & 91.6           & 97.6           & 100.0          & 99.7            \\
                                                  & ncRNA        & 99.7           & 100.0          & 97.6           & 100.0          & 100.0           \\\hline
\multirow{2}{*}{\textit{Macaca mulatta}}          & mRNA         & 85.0           & 94.2           & 99.1           & 100.0          & 100.0           \\
                                                  & ncRNA        & 100.0          & 100.0          & 94.7           & 100.0          & 99.7            \\\hline
\multirow{2}{*}{\textit{Pan troglodytes}}         & mRNA         & 87.1           & 93.9           & 97.7           & 100.0          & 99.8            \\
                                                  & ncRNA        & 99.9           & 100.0          & 88.3           & 99.8           & 99.7            \\\hline
\multirow{2}{*}{\textit{Pongo abelii}}            & mRNA         & 98.0           & 94.4           & 99.9           & 100.0          & 99.9            \\
                                                  & ncRNA        & 100.0          & 100.0          & 98.7           & 99.2           & 99.7            \\\hline
\multirow{2}{*}{\textit{Sus scrofa}}              & mRNA         & 85.1           & 94.9           & 99.3           & 99.9           & 100.0           \\
                                                  & ncRNA        & 98.3           & 98.3           & 76.3           & 99.6           & 99.6            \\ \hline
\multirow{2}{*}{\textit{Xenopus tropicalis}}      & mRNA         & 94.5           & 96.5           & 99.1           & 100.0          & 100.0           \\
                                                  & ncRNA        & 100.0          & 100.0          & 95.0           & 100.0          & 100.0           \\
\hline
\multirow{2}{*}{Average accuracy per class}       & mRNA         & 90.44          & 94.70          & 92.59          & 99.98          & 99.89           \\
                                                  & ncRNA        & 97.75          & 98.61          & 86.98          & 99.63          & 99.77           \\
\textbf{Overall Average Accuracy}                 & \textbf{–}   & \textbf{94.10} & \textbf{96.66} & \textbf{89.78} & \textbf{99.81} & \textbf{99.83}  \\ 
%\hline
\multirow{2}{*}{Standard Deviation}              & mRNA         & 5.28           & 1.46           & 14.71          & 0.04           & 0.14            \\
                                                  & ncRNA        & 3.91           & 2.51           & 16.26          & 0.46           & 0.21            \\
\hline
\end{tabular}
\end{table*}
\endgroup

\begingroup
\setlength{\tabcolsep}{12pt} % Default value: 6pt
\renewcommand{\arraystretch}{1.1} % Default value: 1
\begin{table*}[h]
\centering
\caption{Classification results considering the mRNA, lncRNA and sncRNA classes of sequences from CPC2 dataset \cite{CPC22017}.}
\label{tabelacpc2}
\begin{tabular}{ccccccc} 
\hline
Species                                                                                               & Class of RNA & PLEK           & CPC2           & BASiNET *      & BASiNET        & BASiNETEntropy  \\ 
\hline
\multirow{3}{*}{\textit{Homo sapiens}}                                                                & mRNA         & 97.0           & 95.9           & 80.25          & 100.0          & 99.4            \\
                                                                                                      & lncRNA       & 97.6           & 92.8           & 58             & 100.0          & 99.9            \\
                                                                                                      & sncRNA       & 100.0          & 100.0          & 53.7           & 100.0          & 100.0           \\\hline
\multirow{3}{*}{\textit{Mus musculus}}                                                                & mRNA         & 89.2           & 93.9           & 89.2           & 100.0          & 99.7            \\
                                                                                                      & lncRNA       & 91.7           & 95.0           & 99.8           & 99.9           & 100.0           \\
                                                                                                      & sncRNA       & 100.0          & 100.0          & 72.14          & 99.9           & 99.9            \\\hline
\multirow{3}{*}{\textit{Danio rerio}}                                                                 & mRNA         & 94.4           & 95.5           & 93.3           & 99.5           & 99.7            \\
                                                                                                      & lncRNA       & 79.2           & 88.1           & 99.5           & 98.9           & 99.5            \\
                                                                                                      & sncRNA       & 100.0          & 100.0          & 79.2           & 98.7           & 99.5            \\\hline
\multirow{3}{*}{\begin{tabular}[c]{@{}c@{}}\textit{Arabidopsis }\\\textit{ thaliana}\end{tabular}}    & mRNA         & 63.1           & 99.7           & 96.8           & 99.7           & 99.9            \\
                                                                                                      & lncRNA       & 99.6           & 95.3           & 86.8           & 99.7           & 99.8            \\
                                                                                                      & sncRNA       & 100.0          & 100.0          & 97.3           & 100.0          & 100.0           \\ \hline
\multirow{3}{*}{\begin{tabular}[c]{@{}c@{}}\textit{Caenorhabditis}\\\textit{ elegans}\end{tabular}}   & mRNA         & 53.0           & 96.5           & 77.8           & 100.0          & 100.0           \\
                                                                                                      & lncRNA       & 98.4           & 99.9           & 100            & 99.4           & 99.2            \\
                                                                                                      & sncRNA       & 100.0          & 100.0          & 88.7           & 99.9           & 100.0           \\\hline
\multirow{3}{*}{\begin{tabular}[c]{@{}c@{}}\textit{Drosophila }\\\textit{ melanogaster}\end{tabular}} & mRNA         & 82.8           & 94.6           & 99.9           & 98.5           & 97.3            \\
                                                                                                      & lncRNA       & 87.5           & 91.9           & 84.6           & 97.3           & 99.4            \\
                                                                                                      & sncRNA       & 100.0          & 100.0          & 77.4           & 99.7           & 100.0           \\
\hline
\multirow{2}{*}{Average accuracy per class}                                                           & mRNA         & 79.92          & 96.02          & 89.52          & 99.62          & 99.33           \\
                                                                                                      & ncRNA        & 96.17          & 96.92          & 83.09          & 99.45          & 99.77           \\
\textbf{\textbf{Overall Average Accuracy}}                                                            & \textbf{–}   & \textbf{90.75} & \textbf{96.62} & \textbf{85.23} & \textbf{99.51} & \textbf{99.62}  \\ 
%\hline
\multirow{2}{*}{Standard Deviation}                                                                   & mRNA         & 17.92          & 2.03           & 8.15           & 0.58           & 0.68            \\
                                                                                                      & ncRNA        & 6.67           & 4.18           & 15.13          & 0.81           & 0.29            \\
\hline
\end{tabular}
\end{table*}
\endgroup

\begingroup
\setlength{\tabcolsep}{12pt} % Default value: 6pt
\renewcommand{\arraystretch}{1.2} % Default value: 1
\begin{table*}[h]
\centering
\caption{Classification results considering mRNA and ncRNA sequences for shared species from PLEK \cite{PLEK2014} and CPC2 \cite{CPC22017} datasets. The CPC2 sequences were adopted for the training step and the PLEK sequences were adopted for the classification (test) step.}
\label{cross}
\begin{tabular}{ccccccc}

\hline
Species                                   & Class of RNA      & PLEK           & CPC2           & BASiNET*       & BASiNET        & BASiNETEntropy \\ \hline
\multirow{2}{*}{\textit{Homo sapiens}}     & mRNA        & 90.0           & 94.8           & 88.9           & 99.4           & 99.7            \\
                                           & ncRNA       & 55.0           & 94.1           & 82.9           & 99.2           & 100.0           \\\hline
\multirow{2}{*}{\textit{Danio rerio}}      & mRNA        & 100.0          & 96.1           & 99.7           & 100.0          & 99.7            \\
                                           & ncRNA       & 40.3           & 94.1           & 20.5            & 99.9           & 100.0           \\\hline
\multirow{2}{*}{\textit{Mus musculus}}     & mRNA        & 91.6           & 96.1           & 99.8           & 100            & 99.9            \\
                                           & ncRNA       & 95.8           & 93.1           & 95.0          & 99.9           & 99.7            \\ \hline
\multirow{2}{*}{Average accuracy per class} & mRNA        & 93.85          & 95.67          & 96.13          & 99.80         & 99.76           \\
                                           & ncRNA       & 63.70          & 93.76          & 66.13          & 99.67          & 99.90           \\
\textbf{Overall Average Accuracy}                       & \textbf{--} & \textbf{78.78} & \textbf{94.72} & \textbf{81.13} & \textbf{99.73} & \textbf{99.83}  \\
\multirow{2}{*}{Standard Deviation}             & mRNA        & 4.40           & 0.61           & 6.26           &  0.34           & 0.09            \\
                                           & ncRNA       & 23.48          & 0.47           & 39.9          &  0.40           & 0.14            \\ \hline
\end{tabular}
\end{table*}
\endgroup

The first experiment was performed by considering the PLEK dataset. The proposed method was applied in order to extract the complex network topological measurements (features). Then, the Random Forest \cite{rf} was adopted as a classifier. The competitor methods were performed by considering its default parameters. All methods were performed by considering the 10-fold cross-validation.

Table \ref{tab:plek} presents the results of the classifications considering the classes: mRNAs and ncRNAs. It can be seen that the proposed method showed the highest average accuracy among all the methods compared and lower standards deviations. In particular, it is important to note that BASiNET extracts 2,000 features with the application of the threshold step, leading to a feature space with high dimensionality. When not considering the threshold step, (BASiNET*), it clearly presents a decrease in accuracy, especially considering the ncRNA class. TThe proposed method shows that correctly identify which edges were important to define the topological structure of the network, and as a result, to characterize each class of RNA. Therefore, extracting only 10 features showed remarkable results, contributing to the correct identification of the RNA sequences and also with the reduction of the feature space. Thus, leading to a simpler and more efficient method for feature extraction from RNA sequences and their classification.

The second experiment was performed considering the CPC2 dataset in which the proposed method was applied in the same way as the previous experiment as well as the competitor methods. Table \ref{tabelacpc2} presents the results of the classifications considering the classes mRNAs, lncRNAs and sncRNAs. Again, the BASiNETEntropy showed higher accuracies compared with competitor methods and also with the lowest values of standard deviation, reinforcing the assertiveness and robustness of the proposed method. It is important to point out that BASiNET* again presents a decrease in accuracy, reinforcing the importance of its threshold step when compared with BASiNET.

The third experiment was performed considering a cross-validation between the shared species in PLEK and CPC2 datasets. The RNA classes considered in this experiment were ncRNA and mRNA, so the lncRNA and sncRNA classes from the CPC2 dataset were grouped into a single ncRNA class. In this way, the RNA sequences of each class and species in the CPC2 dataset were adopted for the training step. The respective RNA sequences in PLEK dataset were adopted in the classification (test) step. Therefore, the goal of this experiment was to test the generalization of the methods when trained with sequences from one dataset and tested with sequences from another dataset, considering the same species.

Table \ref{cross} presents the results of the cross-validation between the datasets considering the three shared species: \textit{Homo sapiens},  \textit{Danio rerio} and  \textit{Mus musculus}. It is possible to notice that PLEK method showed a significant decrease in accuracy considering the ncRNA class from \textit{Homo Sapiens} and \textit{Danio rerio} species. On the other hand, the CPC2 method presents similar results to the previous ones, showing its suitability in generalizing the classification of RNA sequences. The BASiNET* method (without the threshold step) presented the lowest accuracy for the ncRNA class considering the \textit{Danio rerio}, showing again the importance of its threshold step when compared to its original version. BASiNET and BASiNETEntropy were the most assertive methods, with a slight superiority of the proposed method in assertiveness and robustness. 

In summary, the results showed that BASiNET* without considering its threshold step had the lowest average accuracy in experiment 1 and 2 and the highest standard deviation. PLEK had the lowest average accuracy in experiment 3, when a cross-validation between data from the adopted datasets was performed, showing low generalization. CPC2 proved stable in all experiments with high accuracy rates and low standard deviation. BASiNET and BASiNETEntropy methods clearly outperform the competitor methods, showing the highest accuracy values and lowest standard deviation in all experiments.

The results show that proposed entropy maximization approach reduces the complexity in terms of dimensionality, extracting only 10 features, while maintaining high accuracy values and low standard deviation. Therefore, the proposed approach proved its efficiency in simplifying the characterization of RNA sequences, maintaining a high assertiveness and robustness in their classification.

% --------------------------------------------------------------------------------------------------
\section{Conclusion}
\label{sec:conclusion}

The classification of biological sequences has become increasingly challenging because of the amount and variety of sequences currently generated \cite{de2021towards}. Traditional methods based on the alignment between sequences require a high computational cost to be performed, being unfeasible for comparing large amounts of data.

This work presents the BASiNETEntropy method as an alignment-free machine learning approach for classifying biological sequences, in particular into different RNA classes. This work is based on a previous method, BASiNET, and presents a significant improvement by eliminating the threshold step that extracts 2,000 features, leading to a high dimensionality in the feature space. Therefore, a filter step based on entropy maximization principle is proposed to select the most representative edges of each class, significantly reducing the number of extracted features.

Two important datasets from the literature were adopted for the assessment of the proposed method, comparing the results with the methods PLEK, CPC2, BASiNET and BASiNET* (removing its threshold step).
The proposed method showed higher accuracies among all competitor methods considering the two adopted datasets. In addition, BASiNETEntropy presented the smallest variations, showing its robustness. Even when cross-validation was performed between the adopted datasets, considering the shared species, the proposed method showed better generalization when classifying the sequences with greater assertiveness among the compared methods. Besides, the proposed method opens the possibility of training and application in other classes of biological sequences.

The BASiNETEntropy method was implemented in open source (R language). The source code, confusion matrices, as well as all the materials for the complete replication of this work are available at \url{https://github.com/fabriciomlopes/BASiNETEntropy}. The software package can also be downloaded at \url{https://cran.r-project.org/web/packages/BASiNETEntropy}.

% --------------------------------------------------------------------------------------------------
\section{Acknowledgments}%
MMB thanks the Universidade Tecnológica Federal do Paraná (UTFPR) for the scholarship (PIBIC 2020/2021). This study was financed in part by the Coordenação de Aperfeiçoamento de Pessoal de Nível Superior - Brasil (CAPES) - Finance Code 001 and the Funda\c{c}\~{a}o Arauc\'{a}ria e do Governo do Estado do Paran\'{a}/SETI (Grant number 035/2019, 138/2021 and NAPI - Bioinform\'{a}tica).

% --------------------------------------------------------------------------------------------------

 % argument is your BibTeX string definitions and bibliography database(s)
%\bibliography{IEEEabrv,references}
%
%\bibliographystyle{IEEEtran}
%\printbibliography%% Referências
\vfill
% Can be used to pull up biographies so that the bottom of the last one
% is flush with the other column.
%\enlargethispage{-5in}

% that's all folks
\end{document}